\documentclass[11pt,twocolappendix,appendixfloats]{emulateapj}
\usepackage{times}
\usepackage[colorlinks=true,citecolor=blue,linkcolor=blue,urlcolor=blue]{hyperref}
\usepackage[usenames]{color}

\usepackage{graphicx}
\usepackage{epstopdf}
\usepackage{url}

\def\src {SGR\,J1745--2900}

\def\ltsima{$\; \buildrel < \over \sim \;$}
\def\simlt{\lower.5ex\hbox{\ltsima}}
\def\gtsima{$\; \buildrel > \over \sim \;$}
\def\simgt{\lower.5ex\hbox{\gtsima}}
\def\ltsima{$\; \buildrel < \over \sim \;$}
\def\lsim{\lower.5ex\hbox{\ltsima}}
\def\loe{\lower.5ex\hbox{\ltsima}}
\def\gtsima{$\; \buildrel > \over \sim \;$}
\def\gsim{\lower.5ex\hbox{\gtsima}}
\def\goe{\lower.5ex\hbox{\gtsima}}
\def\ltsima{$\; \buildrel < \over \sim \;$}
\def\lsim{\lower.5ex\hbox{\ltsima}}
\def\loe{\lower.5ex\hbox{\ltsima}}
\def\gtsima{$\; \buildrel > \over \sim \;$}
\def\gsim{\lower.5ex\hbox{\gtsima}}
\def\goe{\lower.5ex\hbox{\gtsima}}

\def\ergs {erg\,s$^{-1}$}
\def\ergscm2 {erg\,s$^{-1}$cm$^{-2}$}
\def\ss {s\,s$^{-1}$}
\def\cm2 {cm$^{-2}$}

\def\ergs {${\rm erg\, s}^{-1}$}

\begin{document}

\shorttitle{\src: the magnetar mate of Sgr\,A$^*$}
\shortauthors{Rea~et~al.}

\title{A strongly magnetized pulsar within grasp of the \\ Milky Way's supermassive black hole}

\author{N. Rea\altaffilmark{1}, P. Esposito\altaffilmark{2}, J. A. Pons\altaffilmark{3}, R. Turolla\altaffilmark{4,5}, D. F. Torres\altaffilmark{1,6}, 
G. L. Israel\altaffilmark{7}, A. Possenti\altaffilmark{8}, M. Burgay\altaffilmark{8}, D. Vigan\`o\altaffilmark{1,3}, \\ A. Papitto\altaffilmark{1}, R. Perna\altaffilmark{9}, L. Stella\altaffilmark{7},
G. Ponti\altaffilmark{10}, F. K. Baganoff\altaffilmark{11}, D. Haggard\altaffilmark{12}, A. Camero-Arranz\altaffilmark{1}, 
S. Zane\altaffilmark{5}, A. Minter\altaffilmark{13}, S. Mereghetti\altaffilmark{2}, A. Tiengo\altaffilmark{2,14,15}, R. Sch\"odel\altaffilmark{16}, 
M. Feroci\altaffilmark{17}, R. Mignani\altaffilmark{2,5,18}, D. G\"otz\altaffilmark{19} }

\altaffiltext{1}{Institute of Space Sciences (CSIC--IEEC), Faculty of Science, Campus UAB, Torre C5-parell, 2a planta, 08193, Bellaterra (Barcelona), Spain}
\altaffiltext{2}{INAF - IASF, Milano, via E. Bassini 15, I-20133 Milano, Italy}
\altaffiltext{3}{Departament de Fisica Aplicada, Universitat d'Alacant, Ap. Correus 99, 03080 Alacant, Spain}
\altaffiltext{4}{Dipartimento di Fisica e Astronomia, Universit\`a di Padova, via F. Marzolo 8, I-35131 Padova, 
Italy}
\altaffiltext{5}{MSSL-UCL, Holmbury St. Mary, Dorking, Surrey RH5 6NT, UK}
\altaffiltext{6}{ICREA, Barcelona, Spain}
\altaffiltext{7}{INAF - OAR, via Frascati 33, I-00040 Monteporzio Catone, Italy}
\altaffiltext{8}{INAF - OAC, loc. Poggio dei Pini, strada 54, I-09012 Capoterra, Italy}
\altaffiltext{9}{JILA, University of Colorado, Boulder, CO 80309-0440, USA}
\altaffiltext{10}{Max Planck Institute fur Extraterrestriche Physik, D-85748 Garching, Germany}
\altaffiltext{11}{Kavli Institute for Astrophysics and Space Research, Massachusetts Institute of Technology, Cambridge, MA 02139, USA}
\altaffiltext{12}{Center for Interdisciplinary Exploration and Research in Astrophysics, Physics and Astronomy Department, Northwestern University, 2145 Sheridan Rd, Evanston, IL 60208, USA}
\altaffiltext{13}{National Radio Astronomy Observatory, Green Bank, WV 24944, USA}
\altaffiltext{14}{IUSS, Piazza della Vittoria 15, I-27100 Pavia, Italy}
\altaffiltext{15}{INFN, Sezione di Pavia, via A. Bassi 6, I-27100 Pavia, Italy}
\altaffiltext{16}{Instituto de Astrof\'isica de Andaluc\'ia (CSIC), Glorieta de la Astronomia S/N, E-18008 Granada, Spain}
\altaffiltext{17}{INAF -- IAPS, via del Fosso del Cavaliere 100, I-00133 Roma}
\altaffiltext{18}{University of Zielona G\'ora, Lubuska 2, 65-265, Zielona G\'ora, Poland}
\altaffiltext{19}{AIM (CEA/DSM-CNRS-Universit\'e Paris Diderot), Irfu/Service d'Astrophysique, Saclay, F-91191 Gif-sur-Yvette, France.}

\begin{abstract} 
The center of our Galaxy hosts a supermassive black hole, Sagittarius (Sgr) A$^*$. Young, massive 
stars within 0.5 pc of Sgr\,A$^*$ are evidence of an episode of intense star formation near the black hole a few 
Myr ago, which might have left behind a young neutron star traveling deep 
into Sgr\,A$^*$'s gravitational potential. On 2013 April 25, a short X-ray burst was observed from the direction of 
the Galactic center. Thanks to a series of observations with the \emph{Chandra} and the {\em Swift} satellites, we pinpoint the associated magnetar at an angular distance of $2.4\pm 0.3$ arcsec from Sgr\,A$^*$, and refine the source spin period and its derivative ($P=3.7635537(2)$\,s and $\dot{P} = 6.61(4)\times 10^{-12}$\,s s$^{-1}$), confirmed by quasi simultaneous radio observations performed with the Green Bank (GBT) and Parkes antennas, which also constrain a Dispersion Measure of $\rm{DM}=1750 \pm 50$ pc cm$^{-3}$, the highest ever observed for a radio pulsar. We have found that this X-ray source is a young magnetar at $\approx$0.07--2\,pc from Sgr\,A$^*$. Simulations of its possible motion around Sgr\,A$^*$ show that it is likely ($\sim$90\% probability) in a bound orbit around the black hole. The radiation front produced by the past activity from the magnetar passing through the molecular clouds surrounding the Galactic center region, might be responsible for a large fraction of the light echoes observed in the Fe fluorescence features. 
\end{abstract}
 

\section{Introduction}

Among the large variety of Galactic neutron stars, magnetars constitute the most energetic and unpredictable class \citep{mereghetti08}. They are X-ray pulsars rotating at relatively long periods (0.3--12\,s, with $\dot{P}\sim10^{-15}$--$10^{-10}$\ \ss), and persistent X-ray luminosities of $L_{\rm X}\sim10^{33}$--$10^{35}$\ \ergs . They exhibit flaring activity, generally classified as \emph{giant flares} ($\sim10^{46}$--$10^{47}$\,erg emitted in several minutes), \emph{intermediate flares} ($10^{42}$--$10^{45}$\,erg in a few minutes) or  \emph{short X-ray bursts} ($10^{38}$--$10^{40}$\,erg in less than a second), as well as large increases of the persistent flux (outbursts) which can last about 1 yr \citep{rea11}. The powerful emission observed from these objects has been attributed to their exceptionally high magnetic field ($B\sim10^{14}$--$10^{15}$\,G at the star surface), hence the name magnetars \citep{duncan92,thompson93}. 

Analysis of stellar orbits has demonstrated that a supermassive black hole (SMBH) of about $4.3\times10^6$ $M_\odot$ 
resides at the dynamic center of our Galaxy \citep{ghez08,gillessen09}. The black hole is associated with 
Sgr\,A$^*$, a compact source with non-thermal emission in radio, infrared, and X-rays \citep{melia01,baganoff03,genzel10}. 
X-ray fluorescence lines from numerous molecular clouds near the Galactic center have been interpreted as evidence of 
interaction with a light echo tracing a past bright state of Sgr\,A$^*$ \citep{ponti13}.

\begin{figure*}
\includegraphics[width=9cm]{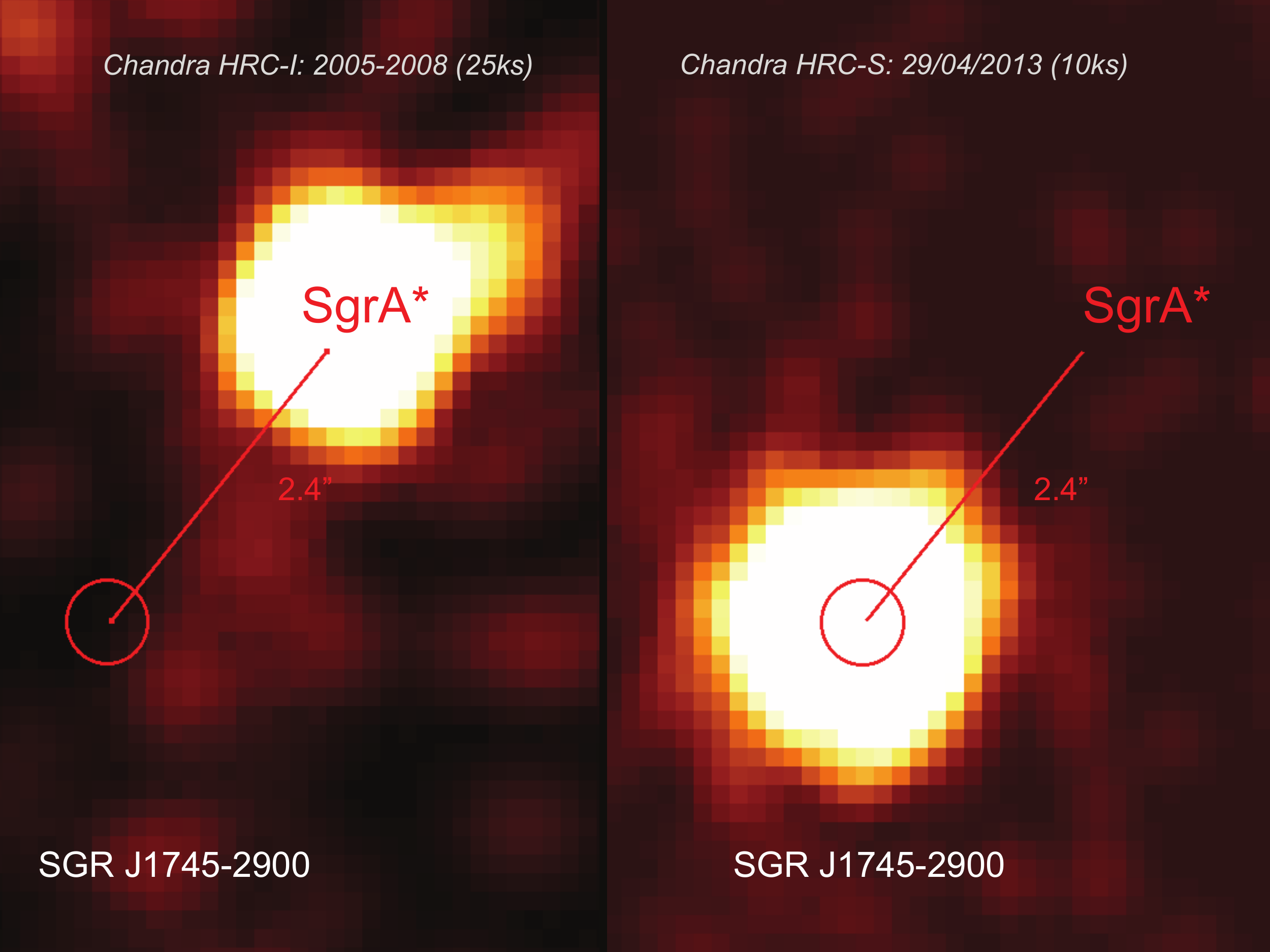}
\includegraphics[width=10cm,height=7.8cm]{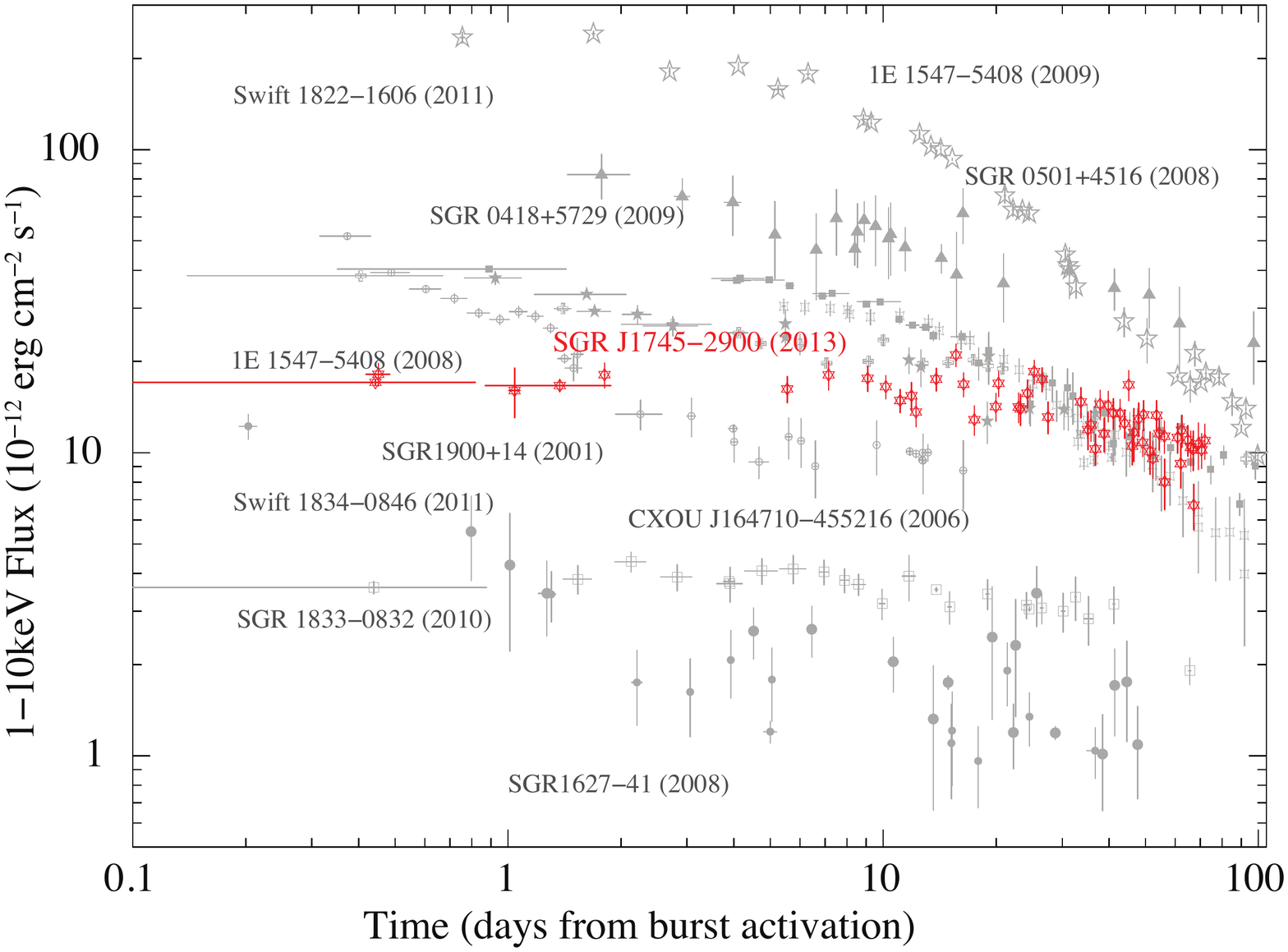}
\caption{\label{hrcfov} {\em Left panel}: {\em Chandra} HRC images of the field of Sgr\,A$^*$ obtained before (co-adding all previous observations performed with the HRC camera) and four days after (on 2013 April 29) the onset of the outburst of SGR\,J1745--2900. {\em Right panel}: \emph{Swift}/XRT long-term flux evolution of \src\ in the 1--10 keV band (red squares) compared to other magnetar outbursts (adapted and updated from \cite{rea11}). \label{figure:xrtlc}}
\end{figure*}

On 2013 April 24, \emph{Swift} detected powerful X-ray emission  from the direction of Sgr\,A$^*$ \citep{degenaar13,kennea13}, initially interpreted as an X-ray flare from 
the SMBH. One day later, a short X-ray burst was observed from a position consistent with that of Sgr\,A$^*$, very 
similar in fluence, duration and spectrum to those commonly observed from magnetars. The magnetar picture received 
further support when \emph{NuSTAR} observed the region, and detected a persistent source with 
periodic modulation at 3.76\,s  \citep{mori13}, a typical value for magnetar spin periods. 


In this Letter we report on X-ray ({\em Chandra} and {\em Swift}) and radio (Robert C. Byrd Green Bank Telescope (GBT) and Parkes Radio Telescope) observations of \src, which point to a 
likely physical connection of the source with Sgr\,A$^*$.

\section{X-ray observations}

\emph{Chandra} \citep{weisskopf03} observed \src\ for the first time on 2013 April 29 with the High Resolution Camera (HRC-S), and three other times in the following three months with the Advanced CCD for Imaging Spectrometer (ACIS-S; see Table\,\ref{cxologs}).  The HRC-S instrument (timing mode with a 0.14" pixel size) observed \src\, for about 10\,ks. The three ACIS-S observations were performed in faint data mode with the 1/8 chip sub-array (the source was positioned in the back-illuminated ACIS-S3 CCD at the 
nominal target position; time resolution 0.441\,s). The data were reduced following standard procedures using the \emph{Chandra} Interactive Analysis of Observations software (CIAO, version 4.5) and the calibration database CALDB 4.5.6. 

After the discovery of \src, \emph{Swift} observed its field almost daily (see also \cite{degenaar13,kennea13}). In this work we use all the observations performed between  TJD 16406 and 16480, taken in both photon counting (PC) and windowed timing (WT) modes. Only the WT observations could be used for the timing analysis of \src\ (Section\,\ref{xtiming}; readout time of 1.7\,ms), owning to the slow read out time of the PC mode (2.5\,s). The data were processed and filtered with standard criteria using Ftools within  HEAsoft software package (v6.12). 

For the ACIS observation, the spectra, ancillary response files and spectral redistribution matrices were created using the CIAO script specextract. For the XRT observations, we used the latest available spectral redistribution matrix in CALDB (v013/v014), while the ancillary response files were generated with xrtmkarf, and they account for different extraction regions, vignetting and PSF corrections. 

We have converted all photon arrival times to the Barycentric Dynamical Time (TDB) system, using the accurate position derived from the \emph{Chandra} observations (see below). 

\subsection{Absolute astrometry}
\label{astrometry}

In the ACIS-S observations a number of sources were detected besides \src. In particular, four known sources present in an X-ray catalog \citep{muno09} were located within $\sim$$30''$ from the SGR and could be employed to refine the absolute astrometry. We used the four source positions, calculated with wavdetect, to register the ACIS-S images on the catalog by optimizing a roto-translation (using the CIAO script reproject\_aspect). The fit yielded an average rms of $\sim$80 mas. The estimated coordinates (taking into account also the accuracy of the reference positions) of \src\ are: $\rm RA =17^h45^m40\fs169$, $\rm Dec= -29^\circ00'29\farcs84$ (J2000.0) with a  95\% confidence level uncertainty radius of $0\farcs3$ (see Fig.\,\ref{hrcfov}).

\begin{table*}
\centering
\caption{Journal of the \emph{Chandra}, Green Bank and Parkes observations.} \label{cxologs}\label{tab:radio}
\begin{tabular}{@{}lcccc}
\hline
\hline
 \multicolumn{5}{c}{{\em Chandra} X-ray observations} )\\
 \hline
Obs.\,ID (Instrument) &  Start date & Count-rate & Exposure & 1--10\,keV Flux \\
 &  (yy-mm-dd hh:mm:ss)  & (counts/s) & (ks) & ($10^{-12}$\ergscm2 ) \\
\hline
14701 (HRC-S)  & 2013-04-29 15:19:50 & 0.087(2) &  9.8 & -- \\
14702 (ACIS-S) & 2013-05-12 10:45:41 &  0.383(5) & 15.1 &  9.2(3)\\
14703 (ACIS-S)  & 2013-06-04 08:45:16 & 0.310(4) & 16.8 &  7.3(2)\\
14946 (ACIS-S) & 2013-07-02 06:57:56 &  0.274(4) & 18.2 & 6.3(2)\\
\hline
\hline
 \multicolumn{5}{c}{Radio observations} )\\
 \hline
 Telescope (Backend)  & Start UT Time &  Frequency  (Bandwidth) & Length
& Flux  \\
& (yy-mm-dd hh:mm:ss) & MHz (MHz) & (ks) &(mJy) \\
\hline
PKS (PDFB4) & 2013-04-27 13:53:04 &  3094 (1024)  & 5.4 & $<$0.02 \\
PKS (PDFB4) & 2013-04-28 15:21:34 &  3094  (1024)  & 1.2 & $<$0.05 \\
GBT (GUPPI) &2013-04-29 07:14:54 &  2000  (800)  & 1.5 & $<$0.02 \\
PKS (PDFB4) & 2013-04-29 16:27:38 & 3094  (1024)  & 5.4 & 0.06 \\
 PKS (PDFB4) &2013-05-01 14:45:01 & 3094  (1024) & 1.2 & 0.1 \\
GBT (GUPPI) &  2013-05-04 11:35:13 & 8900  (800)  & 1.2 & 0.04 \\
\hline
\end{tabular}
\end{table*}

\begin{figure*}
\includegraphics[width=6cm,height=6.2cm]{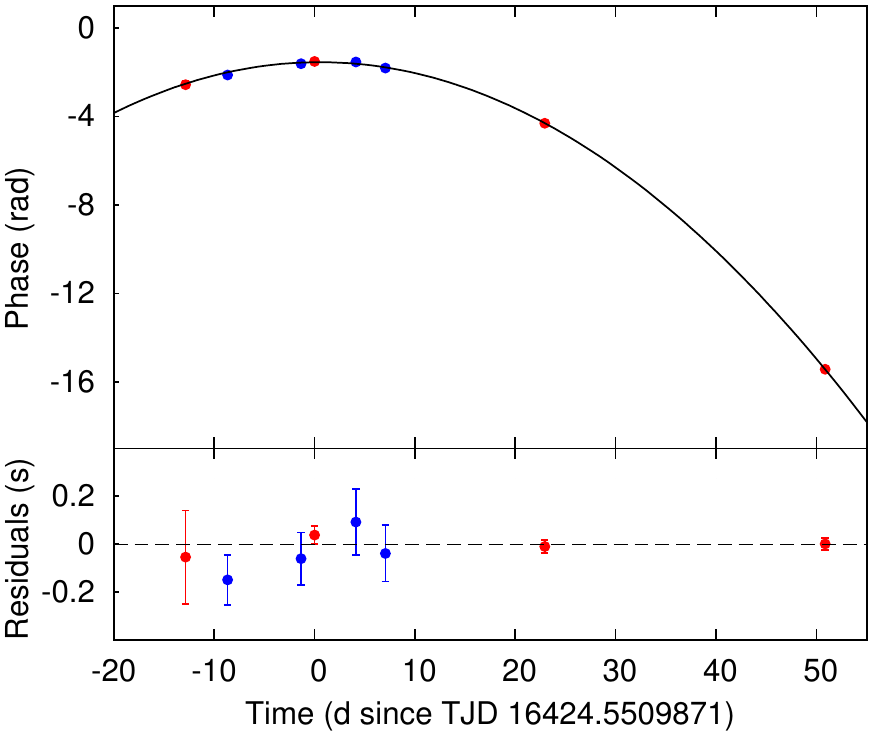}
\hspace{0.2cm}
\includegraphics[width=7cm,height=7.5cm]{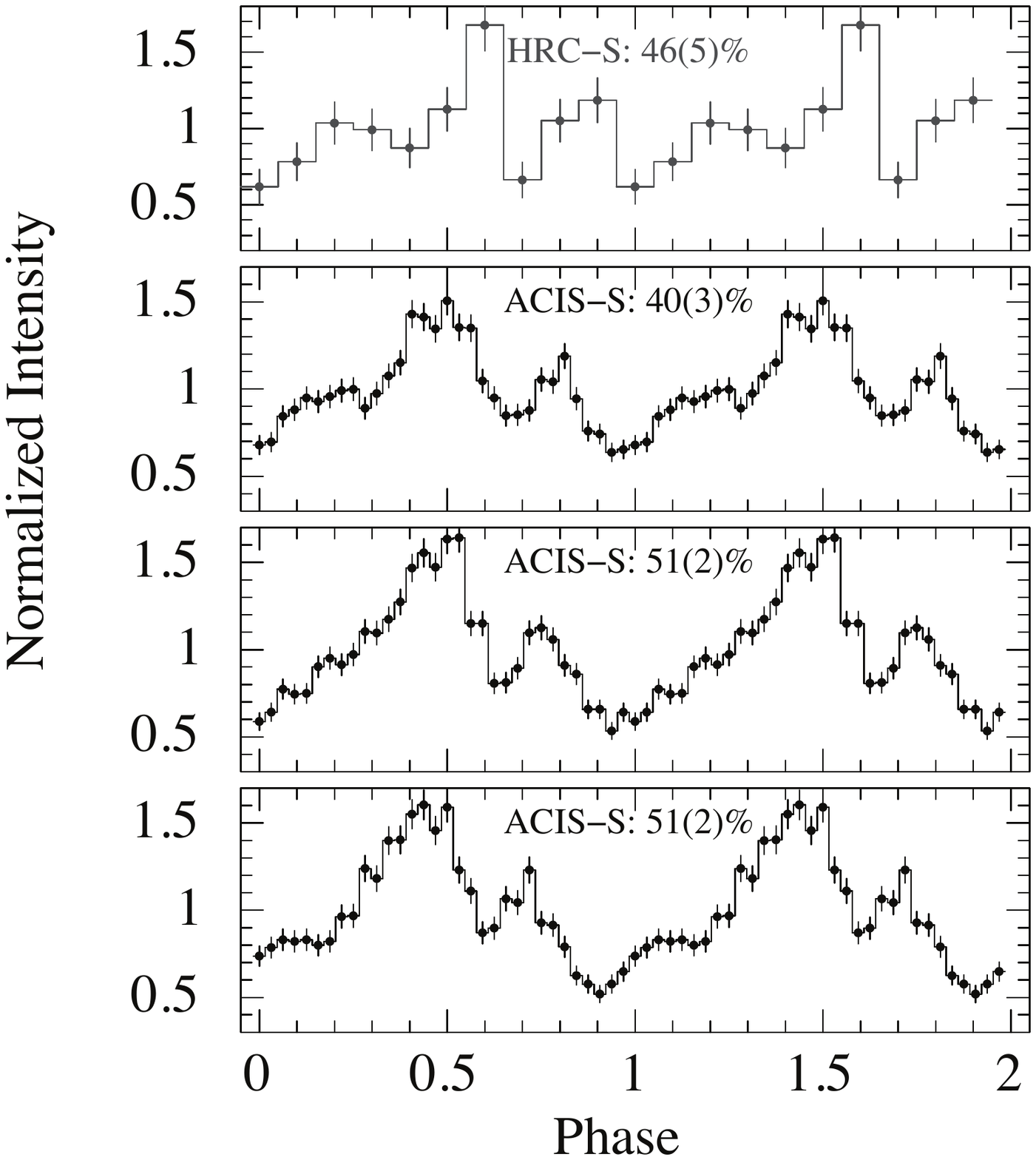}
\hspace{-2cm}
\includegraphics[width=6cm,height=6.5cm]{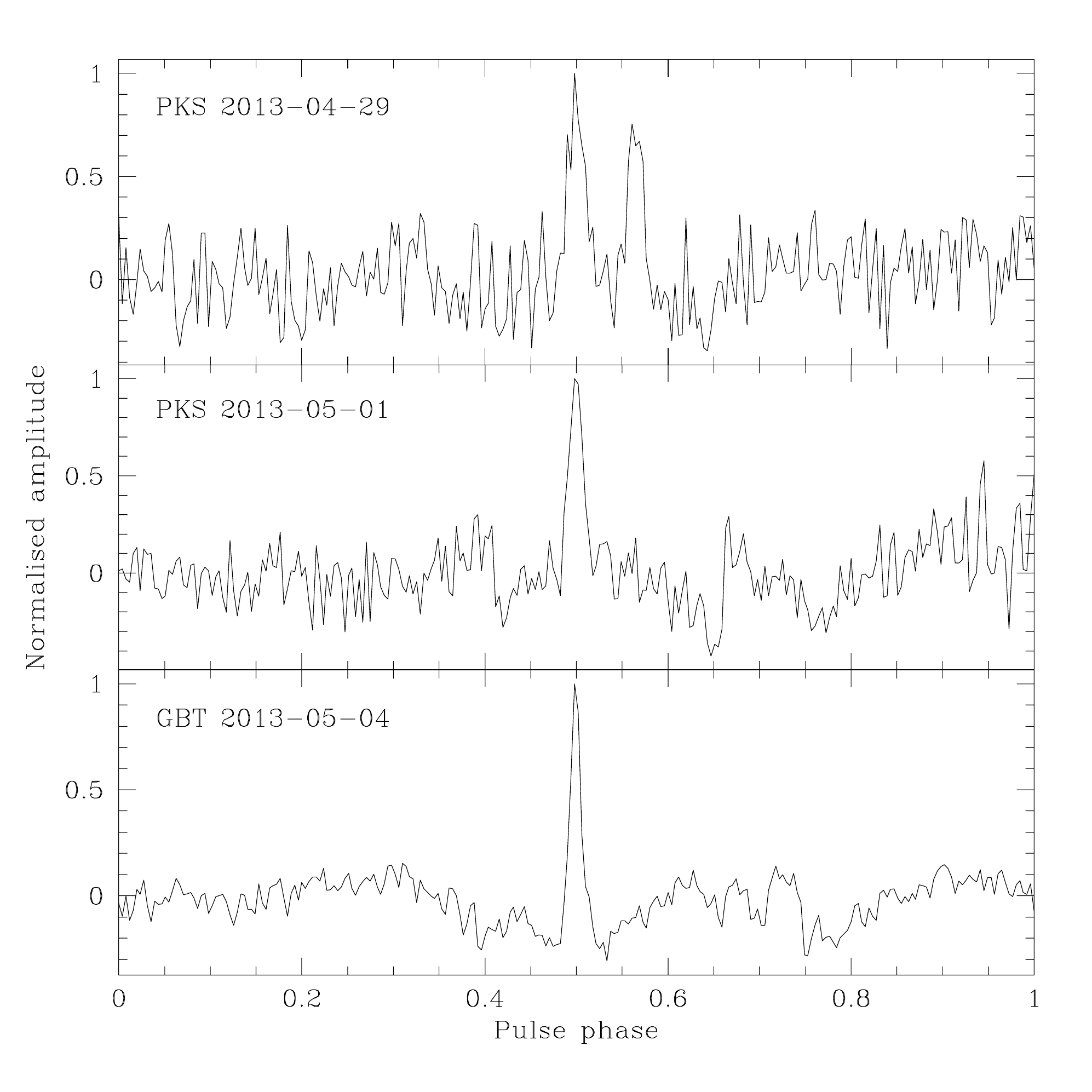}
\caption{\label{fig:timing}{\em Left panel}: evolution of the rotational phase and the residuals with respect to the timing solution reported in the text (\S\ref{xtiming}; in red the {\em Chandra} observations and in blue the {\em Swift} ones). {\em Middle panel}: Pulse profiles of the four {\em Chandra} observations with the relative pulsed fraction (defined as (Max $-$ Min)$/$(Max $+$ Min) of the pulse profile). {\em Right panel}: Pulse profiles of the three radio observations where the magnetar was detected (see \S\ref{radio} and Table\,\ref{cxologs}).
}
\end{figure*}

\subsection{X-ray timing analysis}
\label{xtiming}

Because of the complex shape and variability of the three peaked pulse profile (Fig.\,\ref{fig:timing}),
we decided not to use a pulse template for the timing analysis (which
might artificially affect the phase shift), using instead two parallel
methods: fitting a sinusoid to the profile at the fundamental period,
and fitting the highest peak in all the observations with a
Gaussian. We used 12 phase bins to produce a folded pulse profile. We built up our timing solution from the phases of the
second \emph{Chandra} pointing which has the highest number of
counts. The resulting best-fit period for this \emph{Chandra}
observation was $P=3.763554(2)$ s (at TJD 16424.5509871; TJD = JD -- 2440000.5 days). The accuracy on this measure of the period, 2 $\mu$s, is
enough to coherently phase-connect adjacent observations. 
At each step, we checked that the timing solution was accurate enough
to determine univocally the phase of the following observation. To
this end, we propagated the error affecting the best-fitting
parameters of the solution obtained at any step, to the epoch of the
newly added observation. We never obtained a phase uncertainty larger
than 0.4, which allowed us to tentatively maintain the phase connection. The phase-coherent solution obtained considering the
phases of the best-fit fundamental harmonic component has a period $P=3.7635537(2)$\,s and period derivative $\dot{P} =
6.61(4)\times 10^{-12}$\,s s$^{-1}$ (epoch TJD 16424.5509871), with a reduced chi-squared of 0.85 for 5 degrees of freedom (dof). On the
other hand, when comparing this results with the timing solution derived considering the phases of the best-fit Gaussian to the main peak, we find compatible numbers for the period and the period derivative, but with a marginal evidence for a cubic component (resulting in a possible $|\ddot{P}|=4(3)\times10^{-19}$\,s\,s$^{-2}$).

\begin{figure*}
\centering
\includegraphics[width=.7\textwidth]{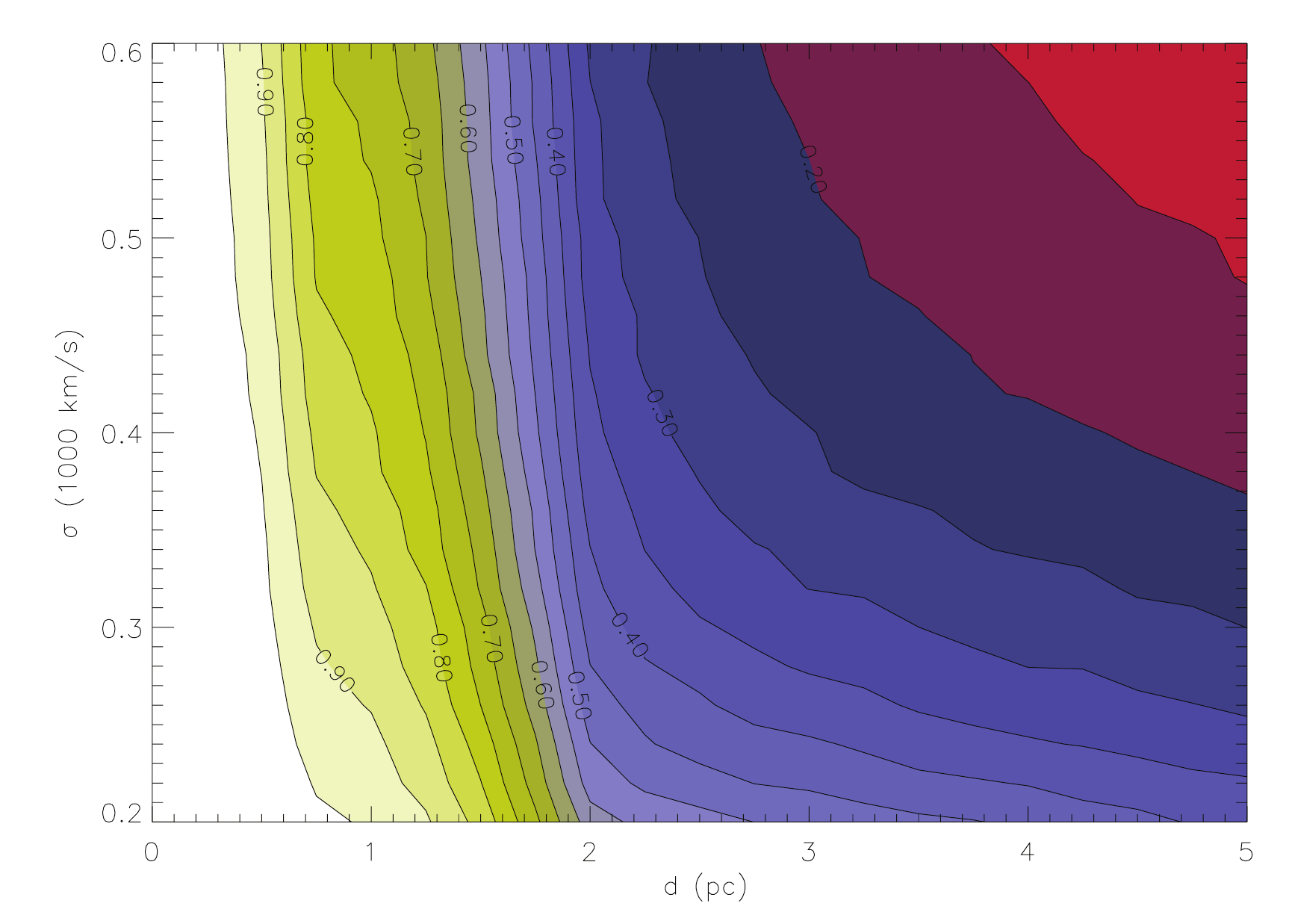}
\caption{\label{orbit} Probability for \src\ to be in a bound orbit around the Galactic center SMBH obtained from Monte Carlo simulations (see text for details).}
\end{figure*}

\subsection{X-ray spectral analysis}
\label{spectra}

In the \emph{Chandra} observations, the pileup in the ACIS-S3 detector was $\sim$5--15$\%$, therefore an annular region with $0.75''$ inner radius (1.5 pixels) and $2''$ outer radius was selected to extract the source spectrum. For the background we used an annular region of maximum radius $8''$ around the source. The point source spectrum was rebinned to have at least 50 counts per energy bin.

We modeled the spectra using the XSPEC v.12.8.0j analysis package. An absorbed single blackbody component (BB) provides a good fit to the data of the three ACIS--S observations ($\chi_r^2\sim1.13$ with 313 dof); the best-fit parameters are: $N_{\mathrm H}= 1.03(4)\times 10^{23}$\,cm$^{-2}$ (abundances and cross-section from \cite{angr} and \cite{bcm98}), with a temperature cooling from $kT=0.95(2)$\,keV, to $kT=0.90(2)$\,keV and then $kT=0.88(2)$\,keV for the three observations, respectively. The blackbody radius remains constant within errors at $R_{\mathrm{BB}}\simeq1.4$\, km (assuming a 8.3\,kpc distance). The 1--10\,keV source flux (see Table\,\ref{cxologs}) corresponds to a luminosity of 2.1, 1.8 and 1.6 $\times10^{35}$\ergs (assuming a 8.3\,kpc distance).

However, we note that the fit residuals are not optimal at higher energies, possibly due to the presence of a non-thermal component. By fitting with a power-law function we also obtain a good modelling ($\chi_r^2\sim1.04$ with 313 dof), with $N_{\mathrm H}= 1.73(4)\times 10^{23}$\,cm$^{-2}$, with photon index of $\Gamma=3.9(1)$, 4.1(1) and 4.2(1) for the three observations. The residuals are slightly better shaped at higher energy. A composition of a blackbody plus power-law model is not statistically required by the data, and does not improve the fit.

The X-ray flux decay  can be modeled with an exponential
function of the form $F(t)=F_\circ\,e^{(-(t-t_\circ)/\tau)}$
($\chi^2_\nu=1.44$ for 80 dof). We fixed $t_\circ$ at the time of the
first burst detected; the resulting best-fit parameters
are $F_\circ=1.72(3)\times10^{-11}$ erg cm$^{-2}$ s$^{-1}$ and
$e$-folding time $\tau=$144(8)\,days. This is a rather slow flux decay as
compared with other outbursts (Fig.\,\ref{figure:xrtlc}).

\section{Radio observations}
\label{radio}

We have also started a monitoring campaign in the radio band using the 64-m Parkes (NSW, Australia) and the 100-m GBT  (WV, USA) radio telescopes to study the magnetar's radio emission \citep{eatough13}. Given the very high electron column density expected in the central regions of the Galaxy we chose to observe at 3.1\,GHz (over a bandwidth of 1024 MHz split into 512 frequency channels) at Parkes, and at 2 and 8.9 GHz (with a bandwidth of 800 MHz split into 2048 frequency channels) at GBT (see Table\,\ref{tab:radio}). At Parkes observations were performed using the ATNF digital filterbank (DFB4; \cite{ferris04}) in search mode, 2-bit sampling the data every 125 $\mu$s, while at GBT the Green Bank Ultimate Pulsar Processing Instrument (GUPPI; \cite{duplain08}) was  used in search mode, 8-bit sampling the data every 64 $\mu$s. For a  faster analysis, GBT data were downsampled in time by a factor of 4 and in frequency by a factor of 8.
Data show that at  2--3\,GHz, the magnetar switched on 
between 2013 April 28 and 29. During the Parkes observation the magnetar 
pulse profile showed a two peaks profile, that evolved into a single one in all subsequent detections (Fig.\,\ref{fig:timing}). From the 10-cm Parkes observations, we obtained a dispersion measure 
$\rm{DM}=1750 \pm 50$~pc~cm$^{-3}$ measured using the {\tt Tempo2} software package (consistent with the value derived by \citet{eatough13,shannon13}).

\section{discussion} 

We performed a series of \emph{Chandra} observations of the Galactic center, determing the position of a new 
transient magnetar in the crowded region of Sgr\,A$^*$ (see Fig.\,\ref{hrcfov}). \src\, is at an angular distance from the position of Sgr\,A$^*$ 
\citep{petrov11} of only $2.4\pm0.3$ arcsec. The timing parameters derived in section \S\ref{xtiming}, imply a surface dipolar magnetic field at the equator 
$B_{\mathrm p} \sim 3.2\times 10^{19}\sqrt{P \dot{P}}\,{\mathrm G}=1.6\times 10^{14}$\,G, a rate of rotational 
energy loss $\dot{E}\sim3.9\times 10^{46} \dot{P}/P^3$\,\ergs $=4.9\times 10^{33}$\ergs , and a characteristic age 
$\tau_c \sim P/(2\dot{P})\sim 9$\,kyr.
 
\begin{figure*} 
\centering
\includegraphics[width=.7\textwidth]{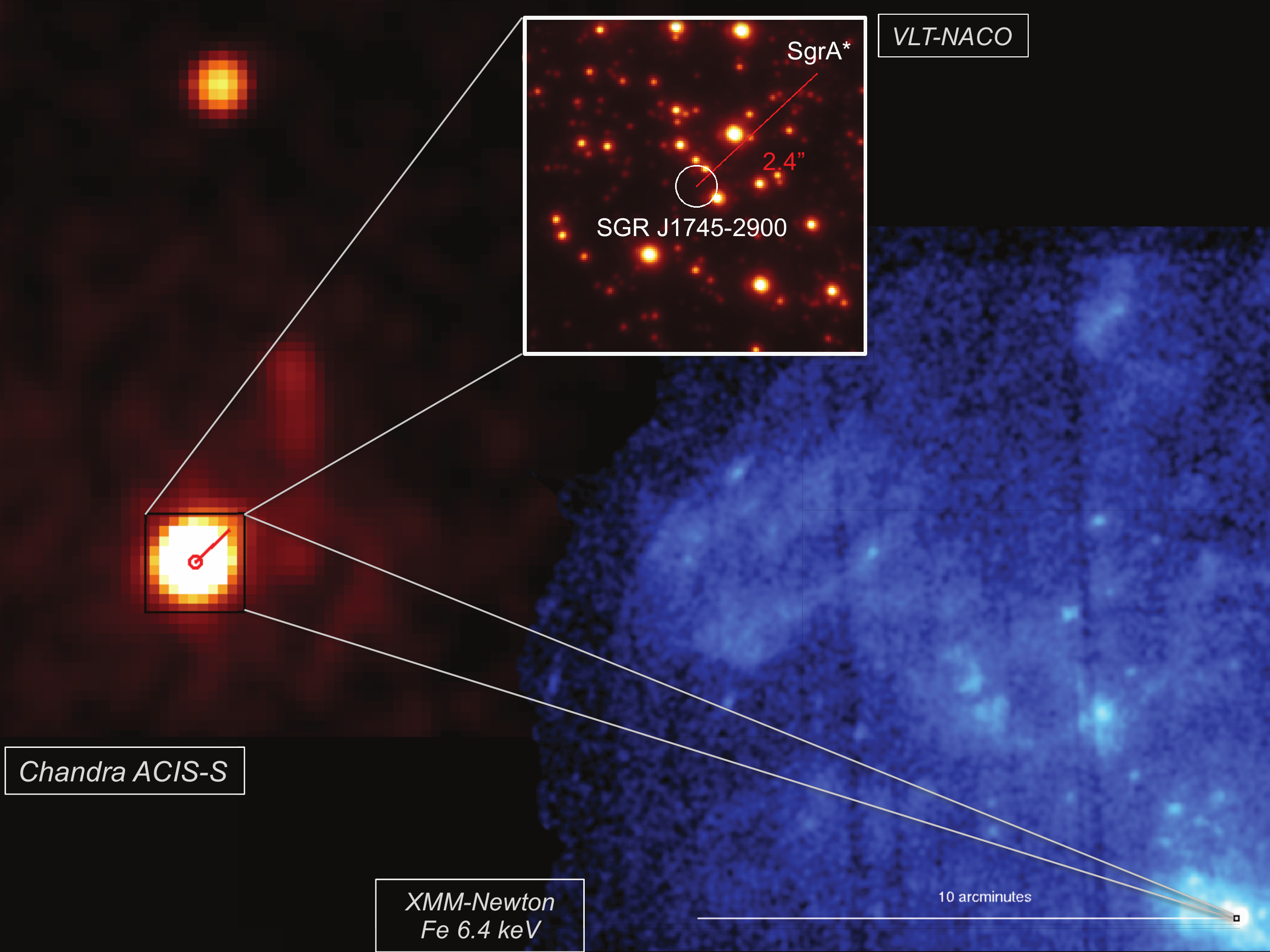}
\caption{\label{field} Multiwavelength view of the field of SGR\,J1745--2900 and Sgr\,A$^*$. The blue image shows the 
XMM-Newton 6.4\,keV Galactic center view \citep{ponti13}, and 
the black square reports a $5''\times5''$ box around the position of the magnetar. The inset shows our first \emph{Chandra} ACIS observation. We also show 
the dense massive star population as observed by VLT/NaCo in the near-infrared K$_{s}$ band \citep{schodel09}.} 
\end{figure*} 

Radio observations performed with Parkes and GBT detected the magnetar radio emission \citep{eatough13}, with the flux and pulse profile variability typical of radio magnetars. The dispersion measure derived from Parkes observations at 10\,cm, $DM=1750\pm50$\,pc\,cm$^{-3}$, is the highest ever measured for a radio pulsar, and implies a distance of $\sim$8.3\,kpc \citep{cordes02}. The relatively high $\dot{E}\sim5\times10^{33}$ \ergs of \src\, compared with the deep limit of its quiescent emission derived by several Ms of {\em Chandra} observations ($\sim10^{32}$ \ergs in the 2--10\,keV range; \citet{muno05}) supports the predictions of the fundamental plane for radio magnetars \citep{rea12}.

\subsection{Association between \src\, and Sgr\,A$^*$}
\label{orbits}

The difference between the column densities of \src\, (see section \S\ref{spectra}) and SgrA$^*$ \citep{baganoff03} translates into $|N_{\mathrm{H,\,J1745}} - 
N_{\mathrm{H,\,SgrA^*}}| \lsim 0.6\times10^{23}$\cm2 \ (at 90\% confidence level). Given that the Central Molecular 
Zone (CMZ) has a particle density $N_{\mathrm{CMZ}}>10^{4}$\,cm$^{3}$ \citep{morris96}, this yields an upper limit on the distance 
between the two sources $d_\mathrm{max}=|N_{\mathrm{H,\,J1745}} - N_{\mathrm{H,\,SgrA^*}}|/N_{\mathrm{CMZ}} \sim$ 2\,pc.

\src\ and Sgr\,A$^*$ relative distance ranges between that implied by their angular separation on the sky, 
$d_{\mathrm{min}}=0.09\pm0.02$\,pc, and $d_{\rm max}\sim 2$\,pc. With an estimated neutron star density in the 
Galactic disc $\approx 3\times10^{-4}$ pc$^{-3}$ \citep{faucher06}, we expect $\sim$ 0.2 neutron stars in a cone of aperture $2\farcs4$ which encompasses the Galactic Center, and reaches out to the rim of the Galaxy (assumed distance from the Sun is 20\,kpc). This estimate comprises all neutron stars born during the Galaxy lifetime, in about 1 Gyr.
The probability of \src\ being a neutron star wandering across the line of 
sight, with an age $\lsim $9\,kyr, and lying at such a small angular distance from Sgr\,A$^{*}$ is then
$\sim$$1.8\times10^{-6}$. This makes the possibility that \src\, is a foreground or background object quite 
unlikely.

On the other hand, estimates of the neutron star population in the Galaxy \citep{freitag06} suggest that there are about 
$2\times10^{4}$ neutron stars within 1\,pc from the Galactic center. Since this is the total population born during 
the entire lifetime of the Galaxy (about a Gyr), a new neutron star is born in the same region every $\sim 10^5$\,yr. 
Even on an observational ground alone, about 80 ordinary radio pulsars \citep{wharton12} are expected to be visible 
with 1\,pc from Sgr\,A$^*$, compatible with us seeing only 
one young radio pulsar with a characteristic age $<$10\,kyr . The magnetar nature of this young radio pulsar so close 
to Sgr\,A$^*$ suggests that there might possibly be as many magnetars as ordinary pulsars in the Galaxy 
\citep{rea10}, and/or the Galactic center region is a favorable place for magnetar formation, possibly because of its 
high density of very massive stars. 

The projected position of \src\ places it within the disk of young and massive stars observed within 0.5\,pc of 
Sgr\,A$^*$ \citep{paumard06,lu09}, which is most likely its birthplace (see Fig.\,\ref{field}). \src\, is probably the end product of one of the young massive stars born during a recent star formation activity within the dense gaseous disk around Sgr\,A$^*$, now accreted into the black hole \citep{levin03}.

The probability of  \src\ being in a bound orbit around Sgr\,A$^*$ depends on the distance and the kick velocity of the neutron star at birth. 
In order to estimate the probability that \src\ is in a bound orbit around Sgr\,A$^*$, we performed numerical 
simulations by integrating the equations of motion in the gravitational potential of the central black hole for a large 
number ($\sim$$10^6$) of stars; Newtonian gravity was assumed throughout. For each orbit the initial radial distance 
$d$ ($0.05\, \mathrm{pc}\leq d\leq 5\,\mathrm{pc}$) was fixed and the star position on the sphere was selected by 
generating two uniform deviates. The modulus of the initial velocity was drawn from a Gaussian distribution with specified 
standard deviation $\sigma$ ($200\,\mathrm{km\,s^{-1}}\leq\sigma\leq 600\, \mathrm{km\,s^{-1}}$); the velocity 
direction was determined by generating two further uniform deviates. Orbits were calculated in the time interval 
$0\leq t\leq T$, where $T=9$ kyr is the estimated age of \src. Stars which at $t=T$ are at a projected radial 
distance of 0.07--0.11\,pc are sorted according to the value of their total energy, $E$. Finally the 
fraction of bound orbits is computed as $N(E<0)/N_ {tot}$. Since a large number of runs are required to explore with sufficiently small uncertainties the $d$ and $\sigma$ ranges, we 
resorted to a 2D model, in which orbits are in the plane of the sky only. We checked that results from the 
2D calculations are in good agreement with those of the complete 3D case for the present purposes. The Monte Carlo simulations are summarized in Fig.\,\ref{orbit}, that 
shows the fraction of bound orbits as a function of $d$ and $\sigma$: if the magnetar was born within 1\,pc of Sgr\,A$^*$ (where most of the massive 
stars are located) the probability of being in a bound orbit around the black hole is $\sim$90\%. For a circular orbit of radius 0.1\,pc, the orbital period would be 1430\,yr, while the 
minimum orbital period would be 500\,yr with the pulsar being now at the apastron of a strongly eccentric orbit \citep{liu12}. Orbital periods for different eccentricities and semi-major axes can reach periods of several kyrs.

\subsection{The possible imprint of \src's past activity on the Galactic center Fe fluorence}
\label{fe}

Studies of the Galactic center environment led to the discovery of Fe K$\alpha$ emission \citep{koyama97} 
from molecular clouds within $\sim$300\,pc from Sgr\,A$^*$ (Fig.\,\ref{field}). The fast variability of these Fe emission features \citep{ponti10,clavel13} supports an 
interpretation in terms of irradiation from one or more radiation fronts passing through the different molecular clouds in the 
Galactic Center, emitted $\sim$100\,yr ago.  The luminosity needed to explain the Fe fluorescence can 
be calculated as $L_{\rm 8keV}\sim 6\times10^{38}\times f$\, erg~s$^{-1}$\citep{sunyaev98}, where $f$ is a parameter 
taking into account flares with duration $\Delta t$ shorter than the cloud light-crossing time: $f \simeq 
(R_{\mathrm{cloud}}/c)/\Delta t$. The total required fluence is then $\approx L_{\mathrm{8\,keV}} \times 
T_{\mathrm{echo}} \sim10^{46}$--$10^{47}$\,erg, where $T_{\rm echo} \approx $1--10 yr is the light crossing time of 
the fastest echoing clouds \citep{clavel13}. This energy could have been emitted by a powerful giant flare (similar to the one observed from SGR\,1806--20; \citet{hurley05}) from \src\, about a century ago. Unfortunately a reliable measure of the soft X-ray flux of the peak of SGR\,1806--20's giant flare is not available \citep{hurley05,inan07}, however for the SGR\,1900+14's giant flare the study of the ionosphere disturbance could give an estimate of its 3--10\,keV flux as a factor of $\sim9$ times the hard X-ray emission level \citep{inan99}. This is also in line with the idea of the softer component being due to the thermal emission from an expanding fireball, which in smaller scale bursts are observed to have a spectrum consistent with a blackbody with kT$\sim$7--10\,keV. The non-thermal hard X-ray emission during the flare peak is instead due to non-thermal processes deriving from the particle acceleration caused by the fireball.  The $>50$\,keV peak flux of SGR\,1806--20's giant flare was estimated to be $\sim$20 \ergscm2 by GEOTAIL measurements (the only instrument providing a good estimate of the first $<$1\,s flare flux; \citet{terasawa05}). Assuming a distance of 15\,kpc \citep{mcclure05}, this translates in a luminosity ($>$50\,keV) of $\sim5\times10^{46}$\ergs . A good guess of the 3--10\,keV emission during the SGR\,1806--20's giant flare would then be about an order of magnitude higher luminosity,
 large enough to be the cause (or a substantial part) of the photon flux responsible for the observed Fe emission. Furthermore, the requirement that a very energetic event from \src\ occurred $\sim$100\, yr ago is compatible with the estimated outburst rate of a magnetar with such properties, which is approximately one every 50--100 yrs  \citep{vigano13,perna11}.

\section{conclusions}
\src\ is the first pulsar discovered at a parsec distance from a supermassive black hole with a non-negligible probability of 
being in a bound orbit. Future measurements of the magnetar proper motion and, possibly, spin-down variability 
(the contribution to the magnetar observed $\dot{P}$ due to the acceleration imparted by Sgr\,A$^*$'s gravitational 
well can be up to $\sim8\times10^{-13}$ \ss ), will be key in observing the effects of the supermassive black hole's gravitational potential on the magnetar's evolution .

\acknowledgments

We thank the {\em Chandra} team, director and Scott Wolk, for the extremely efficient work, and Ryan Shannon for carrying out some of the Parkes observations. NR thanks Giovanni Miniutti and Kevin Hurley for useful discussion. The National Radio Astronomy Observatory is a facility of the National
Science Foundation operated under cooperative agreement by Associated
Universities, Inc. . The Parkes radio telescope is part of the Australia Telescope which is funded by the Commonwealth of Australia for operation as a National Facility managed by CSIRO. We acknowledge support by grants AYA 2012-39303, SGR2009-811, iLINK 2011-0303, AYA 2010-21097-C03-02, Prometeo 2009/103, AYA2010-17631, P08-TIC-4075, INAF 2010 PRIN grant, Chandra Awards GO2-13076X, G03-14060X, GO3-14099X and G03-14121X, and an EU Marie Curie IEF (FP7-PEOPLE-2012-IEF-331095).


\bibliographystyle{apj}

\begin{thebibliography}{39}
\expandafter\ifx\csname natexlab\endcsname\relax\def\natexlab#1{#1}\fi



\bibitem[{{Anders \& Grevesse}(1989){Anders} \& {Grevesse}}]{angr} 
{Anders}, E. \& {Grevesse}, N., 1989,  Geochimica et Cosmochimica Acta 53, 197 


\bibitem[{{Baganoff} {et~al.}(2003){Baganoff}, {Maeda}, {Morris}, {Bautz},
  {Brandt}, {Cui}, {Doty}, {Feigelson}, {Garmire}, {Pravdo}, {Ricker}, \&
  {Townsley}}]{baganoff03}
{Baganoff}, F.~K., {Maeda}, Y., {Morris}, M., et~al. 2003, \apj, 591, 891


\bibitem[{{Balucinska-Church \& McCammon}(1998){Balucinska-Church} \& {McCammon}}]{bcm98} 
{Balucinska-Church}, M. \& {McCammon}, D., 1998, \apj, 496, 1044




\bibitem[Clavel et al.(2013)]{clavel13} Clavel, M., Terrier, R., 
Goldwurm, A., et al.\ 2013, arXiv:1307.3954 


\bibitem[{{Cordes} \& {Lazio}(2002)}]{cordes02}
{Cordes}, J.~M. \& {Lazio}, T.~J.~W. 2002, eprint (astro-ph/0207156)




\bibitem[{{Degenaar} {et~al.}(2013){Degenaar}, {Miller}, {Kennea}, {Reynolds},
  {Gehrels}, \& {Wijnands}}]{degenaar13}
{Degenaar}, N., {Miller}, J.~M., {Kennea}, J., et~al. 2013, \apj, 769, 155


\bibitem[{{Duncan} \& {Thompson}(1992)}]{duncan92}
{Duncan}, R.~C. \& {Thompson}, C. 1992, \apjl, 392, L9


\bibitem[DuPlain et al.(2008)]{duplain08} DuPlain, R., Ransom, 
S., Demorest, P., et al.\ 2008, \procspie, 7019,  


\bibitem[{{Eatough} {et~al.}(2013)}]{eatough13}
{Eatough}, R., et~al., 2013, Atel \#5040


\bibitem[{{Faucher-Gigu{\`e}re} \& {Kaspi}(2006)}]{faucher06}
{Faucher-Gigu{\`e}re}, C.-A. \& {Kaspi}, V.~M. 2006, \apj, 643, 332


\bibitem[Ferris 
\& Saunders(2004)]{ferris04} Ferris, R.~H., \& Saunders, S.~J.\ 2004, Experimental Astronomy, 17, 269 



\bibitem[{{Freitag} {et~al.}(2006){Freitag}, {Amaro-Seoane}, \&
  {Kalogera}}]{freitag06}
{Freitag}, M., {Amaro-Seoane}, P., \& {Kalogera}, V. 2006, \apj, 649, 91

\bibitem[{{Genzel} {et~al.}(2010){Genzel}, {Eisenhauer}, \&
  {Gillessen}}]{genzel10}
{Genzel}, R., {Eisenhauer}, F., \& {Gillessen}, S. 2010, Reviews of Modern
  Physics, 82, 3121

\bibitem[{{Ghez} {et~al.}(2008){Ghez}, {Salim}, {Weinberg}, {Lu}, {Do}, {Dunn},
  {Matthews}, {Morris}, {Yelda}, {Becklin}, {Kremenek}, {Milosavljevic}, \&
  {Naiman}}]{ghez08}
{Ghez}, A.~M., {Salim}, S., {Weinberg}, N.~N., et~al. 2008, \apj, 689, 1044

\bibitem[{{Gillessen} {et~al.}(2009){Gillessen}, {Eisenhauer}, {Trippe},
  {Alexander}, {Genzel}, {Martins}, \& {Ott}}]{gillessen09}
{Gillessen}, S., {Eisenhauer}, F., {Trippe}, S., et~al. 2009, \apj, 692, 1075




\bibitem[Hurley et al.(2005)]{hurley05} Hurley, K., Boggs, 
S.~E., Smith, D.~M., et al.\ 2005, \nat, 434, 1098 


\bibitem[Inan et al.(2007)]{inan07} Inan, U.~S., Lehtinen, 
N.~G., Moore, R.~C., et al.\ 2007, \grl, 34, 8103 


\bibitem[Inan et al.(1999)]{inan99} Inan, U.~S., Lehtinen, 
N.~G., Lev-Tov, S.~J., et al.\ 1999, \grl, 26, 3357 



\bibitem[{{Kennea} {et~al.}(2013){Kennea}, {Burrows}, {Kouveliotou}, {Palmer},
  {G{\"o}{\u g}{\"u}{\c s}}, {Kaneko}, {Evans}, {Degenaar}, {Reynolds},
  {Miller}, {Wijnands}, {Mori}, \& {Gehrels}}]{kennea13}
{Kennea}, J.~A., {Burrows}, D.~N., {Kouveliotou}, C., et~al. 2013, \apjl, 770, L24

\bibitem[{{Koyama} {et~al.}(1997){Koyama}, {Kinugasa}, {Matsuzaki},
  {Nishiuchi}, {Sugizaki}, {Torii}, {Yamauchi}, \& {Aschenbach}}]{koyama97}
{Koyama}, K., {Kinugasa}, K., {Matsuzaki}, K., et~al. 1997, \pasj, 49, L7


\bibitem[Levin 
\& Beloborodov(2003)]{levin03} Levin, Y., \& Beloborodov, A.~M.\ 2003, \apjl, 590, L33 


\bibitem[{{Liu} {et~al.}(2012){Liu}, {Wex}, {Kramer}, {Cordes}, \&
  {Lazio}}]{liu12}
{Liu}, K., {Wex}, N., {Kramer}, M., {Cordes}, J.~M., \& {Lazio}, T.~J.~W. 2012,
  \apj, 747, 1

\bibitem[{{Lu} {et~al.}(2009){Lu}, {Ghez}, {Hornstein}, {Morris}, {Becklin}, \&
  {Matthews}}]{lu09}
{Lu}, J.~R., {Ghez}, A.~M., {Hornstein}, S.~D., et~al.  2009, \apj, 690, 1463

\bibitem[{{Melia} \& {Falcke}(2001)}]{melia01}
{Melia}, F. \& {Falcke}, H. 2001, \araa, 39, 309


\bibitem[McClure-Griffiths 
\& Gaensler(2005)]{mcclure05} McClure-Griffiths, N.~M., \& Gaensler, B.~M.\ 2005, \apjl, 630, L161 

\bibitem[{{Mereghetti}(2008)}]{mereghetti08}
{Mereghetti}, S. 2008, \aapr, 15, 225

\bibitem[{{Mori} {et~al.}(2013){Mori}, {Gotthelf}, {Zhang}, {An}, {Baganoff},
  {Barri{\`e}re}, {Beloborodov}, {Boggs}, {Christensen}, {Craig}, {Dufour},
  {Grefenstette}, {Hailey}, {Harrison}, {Hong}, {Kaspi}, {Kennea}, {Madsen},
  {Markwardt}, {Nynka}, {Stern}, {Tomsick}, \& {Zhang}}]{mori13}
{Mori}, K., {Gotthelf}, E.~V., {Zhang}, S., et~al.  2013, \apjl, 770, L23


\bibitem[Morris \& Serabyn(1996)]{morris96} Morris, M., \& Serabyn, E.\ 1996, \araa, 34, 645 



\bibitem[Muno et al.(2005)]{muno05} Muno, M.~P., Pfahl, E., 
Baganoff, F.~K., et al.\ 2005, \apjl, 622, L113 



\bibitem[{{Muno} {et~al.}(2009){Muno}, {Bauer}, {Baganoff}, {Bandyopadhyay},
  {Bower}, {Brandt}, {Broos}, {Cotera}, {Eikenberry}, {Garmire}, {Hyman},
  {Kassim}, {Lang}, {Lazio}, {Law}, {Mauerhan}, {Morris}, {Nagata},
  {Nishiyama}, {Park}, {Ram{\`i}rez}, {Stolovy}, {Wijnands}, {Wang}, {Wang}, \&
  {Yusef-Zadeh}}]{muno09}
{Muno}, M.~P., {Bauer}, F.~E., {Baganoff}, F.~K., et~al. 2009, \apjs,
  181, 110


\bibitem[Paumard et al.(2006)]{paumard06} Paumard, T., Genzel, 
R., Martins, F., et al.\ 2006, \apj, 643, 1011 



\bibitem[{{Perna} \& {Pons}(2011)}]{perna11}
{Perna}, R. \& {Pons}, J.~A. 2011, \apjl, 727, L51

\bibitem[{{Petrov} {et~al.}(2011){Petrov}, {Kovalev}, {Fomalont}, \&
  {Gordon}}]{petrov11}
{Petrov}, L., {Kovalev}, Y.~Y., {Fomalont}, E.~B., \& {Gordon}, D. 2011, \aj,
  142, 35



\bibitem[{{Ponti} {et~al.}(2013){Ponti}, {Morris}, {Terrier}, \&
  {Goldwurm}}]{ponti13}
{Ponti}, G., {Morris}, M.~R., {Terrier}, R., \& {Goldwurm}, A. 2013, in
  Astrophysics and Space Science Proceedings, Vol.~34, Cosmic Rays in
  Star-Forming Environments, ed. D.~F. Torres \& O.~Reimer (Springer,
  Heidelberg), 331--370

\bibitem[{{Ponti} {et~al.}(2010){Ponti}, {Terrier}, {Goldwurm}, {Belanger}, \&
  {Trap}}]{ponti10}
{Ponti}, G., {Terrier}, R., {Goldwurm}, A., {Belanger}, G., \& {Trap}, G. 2010,
  \apj, 714, 732

\bibitem[{Rea \& Esposito(2011)}]{rea11}
Rea, N. \& Esposito, P. 2011, in High-Energy Emission from Pulsars and their
  Systems, ed. D.~F. Torres \& N.~Rea, Astrophysics and Space Science
  Proceedings (Springer, Heidelberg), 247--273

\bibitem[{{Rea} {et~al.}(2010){Rea}, {Esposito}, {Turolla}, {Israel}, {Zane},
  {Stella}, {Mereghetti}, {Tiengo}, {G{\"o}tz}, {G{\"o}{\u g}{\"u}{\c s}}, \&
  {Kouveliotou}}]{rea10}
{Rea}, N., {Esposito}, P., {Turolla}, R., et~al. 2010, Science, 330, 944


\bibitem[Rea et al.(2012)]{rea12} Rea, N., Pons, J.~A., 
Torres, D.~F., \& Turolla, R.\ 2012, \apjl, 748, L12 


\bibitem[{{Sch{\"o}del} {et~al.}(2009){Sch{\"o}del}, {Merritt}, \&
  {Eckart}}]{schodel09}
{Sch{\"o}del}, R., {Merritt}, D., \& {Eckart}, A. 2009, \aap, 502, 91

\bibitem[{{Shannon} \& {Johnston}(2013)}]{shannon13}
{Shannon}, R.~M. \& {Johnston}, S. 2013, \mnras, submitted (eprint:
  astroph/1305.3036)


\bibitem[{{Sunyaev} \& {Churazov}(1998)}]{sunyaev98}
{Sunyaev}, R. \& {Churazov}, E. 1998, \mnras, 297, 1279


\bibitem[Terasawa et al.(2005)]{terasawa05} Terasawa, T., Tanaka, 
Y.~T., Takei, Y., et al.\ 2005, \nat, 434, 1110 



\bibitem[{{Thompson} \& {Duncan}(1993)}]{thompson93}
{Thompson}, C. \& {Duncan}, R.~C. 1993, \apj, 408, 194




\bibitem[{{Vigan{\`o}} {et~al.}(2013){Vigan{\`o}}, {Rea}, {Pons}, {Perna}, {Aguilera} \&
  {Miralles}}]{vigano13}
{Vigan{\`o}}, D., {Rea}, N., {Pons}, J.~A., {Perna}, R., {Aguilera}, D., \& {Miralles}, J.~A. 2013,  \mnras, 434, 123 

\bibitem[{{Weisskopf} {et~al.}(2003){Weisskopf}, {Aldcroft}, {Bautz},
  {Cameron}, {Dewey}, {Drake}, {Grant}, {Marshall}, \& {Murray}}]{weisskopf03}
{Weisskopf}, M.~C., {Aldcroft}, T.~L., {Bautz}, M.,
  et~al. 2003, Experimental Astronomy, 16, 1

\bibitem[{{Wharton} {et~al.}(2012){Wharton}, {Chatterjee}, {Cordes}, {Deneva},
  \& {Lazio}}]{wharton12}
{Wharton}, R.~S., {Chatterjee}, S., {Cordes}, J.~M., {Deneva}, J.~S., \&
  {Lazio}, T.~J.~W. 2012, \apj, 753, 108

\end{thebibliography}

\end{document}